\documentclass[sigconf,nonacm,screen]{acmart}
\makeatletter
\def\@affiliationfont{\small\normalfont}
\makeatother
\usepackage{tikz}
\usetikzlibrary{positioning,arrows.meta,shapes.geometric,shadows}
\usepackage{pgfplots}
\pgfplotsset{compat=1.17}
\author{Shenhua Gu}
\authornote{Corresponding author.}
\email{kevin.gu@unity.cn}
\affiliation{\institution{Tuanjie Engine, Unity China}\city{Shanghai}\country{China}}
\author{Hongqiang Zhu}
\email{hongqiang.zhu@unity.cn}
\affiliation{\institution{Tuanjie Engine, Unity China}\city{Shanghai}\country{China}}
\author{Fan Zhang}
\email{fan.zhang@unity.cn}
\affiliation{\institution{Tuanjie Engine, Unity China}\city{Shanghai}\country{China}}
\author{Jinming Zhang}
\email{jinming.zhang@unity.cn}
\affiliation{\institution{Tuanjie Engine, Unity China}\city{Shanghai}\country{China}}
\author{Hao Chen}
\email{hao.chen@unity.cn}
\affiliation{\institution{Tuanjie Engine, Unity China}\city{Shanghai}\country{China}}

\begin{document}

\title{Unity Insight: A Production Code--Asset Index for LLM Coding Agents in Unity Projects}

\begin{abstract}
LLM coding agents increasingly operate inside game-engine repositories, where application logic is inseparable from serialized assets: a single gameplay change may span C\# scripts, prefabs, scenes, and ScriptableObjects wired together by Unity GUIDs. The retrieval tools agents carry today---shell utilities and code-only indexes---cannot answer basic cross-file questions, because these relationships live in \texttt{.meta} files and YAML assets rather than in code. We present Unity Insight, to our knowledge the first persistent, LLM-facing, agent-integrated cross-file code--asset index for Unity projects, shipping in production with Tuanjie Codely, the agent CLI of Tuanjie Engine, since its public launch on 2026-07-28. In a paired experiment---28 project-specific questions on two Unity games, same model and harness, one run per arm per question---the index-backed agent spent 53\% fewer tokens and 52\% less wall-clock time than a general-purpose exploration agent (exact paired sign tests, $p{<}0.004$), using only its typed index-query tools.
\end{abstract}

\maketitle

\section{Introduction}\label{sec:intro}

LLM coding agents have crossed from single-file edits to whole-repository engineering. In mainstream repositories this works because the relevant structure lives in source code, where text search and language servers~\cite{lsp} give agents reliable bearings. Game-engine repositories break this assumption. In a Unity project, a script is referenced by prefabs and scenes serialized as YAML and linked through GUIDs recorded in \texttt{.meta} files; gameplay data lives in ScriptableObjects; binary assets leave no textual footprint beyond a GUID. The questions an agent routinely must answer---\emph{which prefabs instantiate this script?} \emph{which scene consumes this texture?}---have no answer in the C\# code alone, and text search over serialized YAML returns opaque identifiers that the agent must chain manually, page by page, at high token cost.

We present \emph{Unity Insight}, a persistent cross-file code--asset index for LLM coding agents. Unity Insight is, to our knowledge, the first persistent, LLM-facing, agent-integrated cross-file code--asset index for Unity projects, shipping with the Codely agent~\cite{codely-cli} since its public launch on 2026-07-28~\cite{insight-docs}. This report introduces the system and publishes the first quantitative evidence we are aware of on Unity-aware hybrid indexing for LLM coding agents.

\section{Unity Insight}\label{sec:system}

\begin{figure}[t]
\centering
\begin{tikzpicture}[x=1mm, y=1mm,
  font=\scriptsize,
  zone/.style={rounded corners=2mm, draw=black!12},
  card/.style={draw=black!55, rounded corners=1.5mm, align=center, inner sep=1.5mm,
               minimum height=12mm, fill=#1, drop shadow={shadow xshift=0.3mm, shadow yshift=-0.35mm, opacity=0.18}},
  chip/.style={draw=blue!45!black, rounded corners=2.2mm, inner sep=1.4mm, font=\tiny\ttfamily, fill=blue!8},
  gn/.style={circle, draw=black!60, fill=#1, minimum size=10.5mm, inner sep=0pt, font=\tiny},
  arr/.style={-{Latex[length=2mm]}, black!70, thick},
  biarr/.style={{Latex[length=2mm]}-{Latex[length=2mm]}, black!70, thick},
  ztitle/.style={anchor=west, font=\tiny\bfseries, text=black!60}
]
\fill[zone, gray!9] (0,47) rectangle (84,64);
\node[ztitle] at (2,61.5) {Unity project};
\node[card=blue!14] at (14,53.5)
  {\tikz{\draw[fill=blue!22,draw=black!65,line width=0.25mm] (0,0)--(0,4mm)--(2.2mm,4mm)--(3mm,3.2mm)--(3mm,0)--cycle;
         \draw[fill=blue!35,draw=black!65,line width=0.25mm] (2.2mm,4mm)--(2.2mm,3.2mm)--(3mm,3.2mm)--cycle;
         \draw[black!65,line width=0.2mm] (0.6mm,3.2mm)--(1.6mm,3.2mm);
         \draw[black!65,line width=0.2mm] (0.6mm,2.5mm)--(2.4mm,2.5mm);
         \draw[black!65,line width=0.2mm] (0.6mm,1.8mm)--(2.4mm,1.8mm);}\\
   \textbf{C\# scripts}};
\node[card=green!14] at (42,53.5)
  {\tikz{\draw[fill=green!30,draw=black!65,line width=0.25mm] (0,1.6mm)--(1.3mm,2.3mm)--(2.6mm,1.6mm)--(1.3mm,0.9mm)--cycle;
          \draw[fill=green!16,draw=black!65,line width=0.25mm] (0,1.6mm)--(1.3mm,0.9mm)--(1.3mm,-0.7mm)--(0,0mm)--cycle;
          \draw[fill=green!24,draw=black!65,line width=0.25mm] (2.6mm,1.6mm)--(1.3mm,0.9mm)--(1.3mm,-0.7mm)--(2.6mm,0mm)--cycle;}\\
   \textbf{prefabs, scenes,}\\\textbf{ScriptableObjects}};
\node[card=orange!20] at (70,53.5)
  {\tikz{\draw[fill=orange!30,draw=black!65,line width=0.25mm,rounded corners=0.5mm,rotate=-25] (0,0) rectangle (3.4mm,1.9mm);
          \draw[fill=white,draw=black!65,line width=0.2mm,rotate=-25] (0.8mm,0.95mm) circle (0.4mm);}\\
   \textbf{\texttt{.meta} files}\\ \tiny (GUIDs)};
\draw[arr] (42,47) -- (42,43);
\node[anchor=west, font=\tiny] at (44,45) {crawl \& extract};
\fill[zone, teal!9] (0,12) rectangle (84,41);
\node[ztitle, text=teal!50!black] at (2,39.3) {Unity Insight index \tiny(off by default, read-only)};
\node[cylinder, shape border rotate=90, aspect=0.35, draw=black!55, fill=teal!18,
      minimum width=16mm, minimum height=11mm,
      drop shadow={shadow xshift=0.3mm, shadow yshift=-0.35mm, opacity=0.18}] at (14,29.5) {\tiny\bfseries SQLite};
\node[font=\tiny, text=black!60] at (14,21.8) {persistent store};
\node[anchor=west, font=\tiny\bfseries, text=teal!50!black] at (31,36.2) {fused code--asset graph};
\node[gn=blue!18]  (s) at (34,29) {script};
\node[gn=green!18] (p) at (56,29) {prefab};
\node[gn=orange!26] (t) at (76,29) {texture};
\draw[dashed, black!40] (22,29.5) -- (28.8,29);
\draw[arr, blue!55!black] (s) -- node[font=\tiny, above=0.3] {used by} (p);
\draw[arr, orange!75!black] (p) -- node[font=\tiny, above=0.3] {uses} (t);
\node[chip] at (9,16)  {vfs\_ls};
\node[chip] at (25.5,16) {vfs\_glob};
\node[chip] at (42,16)  {vfs\_grep};
\node[chip] at (58.5,16) {vfs\_read};
\node[chip] at (75,16) {vfs\_refs};
\draw[biarr] (42,12) -- (42,7.5);
\node[anchor=west, font=\tiny] at (44,9.75) {queries \& results, in the loop};
\node[card=blue!12, rounded corners=1.5mm, minimum height=7.5mm, minimum width=42mm] at (42,3.75)
  {\textbf{LLM coding agent}};
\end{tikzpicture}
\caption{Unity Insight architecture. The crawler fuses C\# sources and serialized assets into a persistent, read-only SQLite code--asset graph; the agent queries the graph in the loop through five typed tools (\texttt{vfs\_ls}, \texttt{vfs\_glob}, \texttt{vfs\_grep}, \texttt{vfs\_read}, \texttt{vfs\_refs}) instead of chaining grep over GUIDs manually.}
\label{fig:arch}
\end{figure}
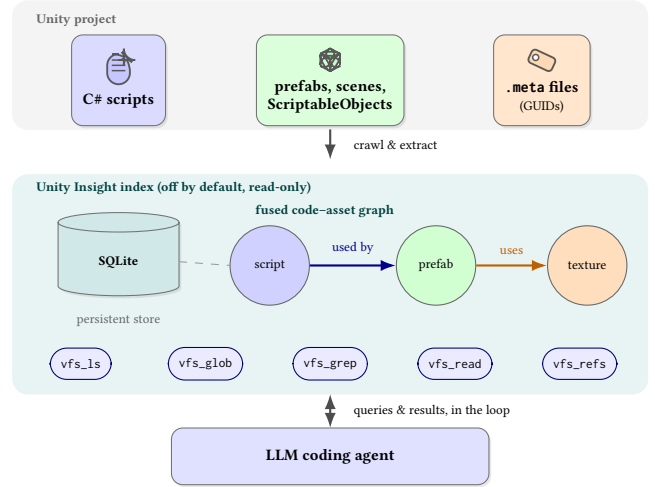

Figure~\ref{fig:arch} shows the architecture. The crawler parses C\# sources and serialized assets---scenes, prefabs, ScriptableObjects, and importer \texttt{.meta} files---resolving GUID$\leftrightarrow$path mappings, component--script semantic bindings, nested-prefab instance chains, and asset-to-asset references into a persistent, read-only SQLite store. The query layer exposes five typed tools to the agent: directory-style listing, path globbing, content search, path-addressed reads, and bidirectional reference traversal (e.g., from a script to its instantiating prefabs, or from a texture to its consumers), designed for token economics: compact, schema-stable responses instead of verbose grep output. The index builds on demand, refreshes incrementally, is off by default, and never mutates project files. Relationship extraction uses typed interfaces where available and regex-based extraction where not; known blind spots (e.g., string-based \texttt{Resources.Load} dynamic loads) are disclosed rather than silently missed. The index is a static snapshot: it deliberately does not capture runtime or editor state.

\section{Experimental Study}\label{sec:study}

\begin{table}[t]
\centering
\caption{Paired cost comparison over 28 project-specific questions (one run per arm per question; same model and harness). Reduction is (explore $-$ unity-insight)/explore on arm totals.}
\label{tab:results}
\footnotesize
\begin{tabular}{@{}lrrr@{}}
\toprule
 & \textbf{explore} & \textbf{unity-insight} & \textbf{reduction} \\
\midrule
Total tokens & 5,000,776 & 2,332,567 & 53.4\% \\
Model output tokens & 196,926 & 111,226 & 43.5\% \\
Wall clock (s) & 3,363 & 1,626 & 51.6\% \\
Tool calls & 509 & 375 & 26.3\% \\
Model rounds & 186 & 145 & 22.0\% \\
\bottomrule
\end{tabular}
\end{table}

\textbf{Setup.} We ran a paired batch study inside the production agent. The arms are two production subagents of the Codely CLI: \texttt{explore}, the general-purpose read-only exploration agent (globbing, text search, file reading, directory listing, and shell access), and \texttt{unity-insight}, whose query surface is the index's typed tools. Subjects are two Unity games: \emph{RedRunner}~\cite{redrunner}, an MIT-licensed 2D platformer, and an internal 3D kart-style racing project. The 28 project-specific questions span asset- and scene-reference chains (\emph{which scene consumes this material? which animation clip does this prefab depend on?}), script discovery, scene-hierarchy and prefab-instance lookups, and multi-artifact impact analysis. Both arms used the same model configuration and the same harness, one run per question per arm; the harness rebuilt the index ahead of each \texttt{unity-insight} session, outside the measured window. Metrics are session-level: total tokens (input including cache, plus output and thinking), model output tokens, agent-loop wall clock, tool calls, and model rounds.

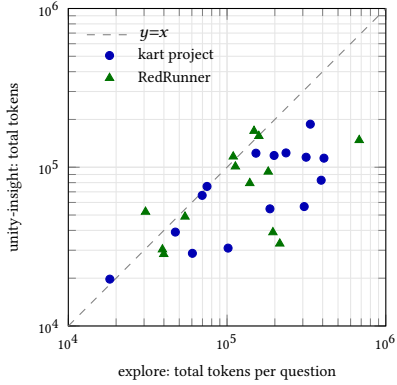
\begin{figure}[t]
\centering
\begin{tikzpicture}
\begin{axis}[
  scale only axis, width=42mm, height=42mm,
  xmode=log, ymode=log,
  xmin=1e4, xmax=1e6, ymin=1e4, ymax=1e6,
  xlabel={\scriptsize explore: total tokens per question},
  ylabel={\scriptsize unity-insight: total tokens},
  grid=both, grid style={gray!20},
  tick label style={font=\tiny},
  legend style={font=\scriptsize, at={(0.03,0.97)}, anchor=north west, draw=none, fill=none, row sep=-1pt},
  legend cell align=left
]
\addplot[gray, dashed, domain=10000:1000000, samples=2, forget plot] {x};
\addlegendimage{gray, dashed}
\addlegendentry{$y{=}x$}
\addplot[only marks, mark=*, mark size=1.5pt, color=blue!70!black] table[x=tokE, y=tokI, col sep=comma] {pairs_kart.csv};
\addlegendentry{kart project}
\addplot[only marks, mark=triangle*, mark size=1.8pt, color=green!45!black] table[x=tokE, y=tokI, col sep=comma] {pairs_red.csv};
\addlegendentry{RedRunner}
\end{axis}
\end{tikzpicture}
\caption{Per-question total tokens (session-level, cache-inclusive; log--log, equal scale---the diagonal $y{=}x$ is a 45-degree reference). Points below the diagonal are questions on which the index-backed arm used fewer tokens: 23/28 overall.}
\label{fig:scatter}
\end{figure}

\textbf{Results.} Table~\ref{tab:results} shows arm totals: the index-backed arm used 53\% fewer tokens and 52\% less wall-clock time. Figure~\ref{fig:scatter} shows the effect is consistent across questions rather than outlier-driven---median per-question ratios are 0.59 for tokens and 0.53 for wall clock---and exact two-sided binomial sign tests confirm it (ties discarded): fewer tokens on 23/28 questions ($p{=}0.0009$), less output on 24/28 ($p{=}0.0002$), less wall time on 22/28 ($p{=}0.0037$), and fewer tool calls on 21/26 ($p{=}0.0025$; two ties). Tool behavior separates cleanly: all 375 of the index arm's calls went through its typed query tools, with zero fallback to raw file I/O, versus 509 file-system and shell calls for the exploration arm (one exploration session even exceeded the agent's context window; we include the triggered 41.8K-token history-compression pass in that arm's totals). We report cost metrics only; mechanical correctness grading enters the controlled study.

\textbf{Example.} One RedRunner question asks: \emph{``Which animation clip does the Chest Big prefab depend on?''} Both arms converged on the same answer---the clip \emph{Open} at \texttt{Assets/Animations/Chest Big/Open.anim}. The index arm reached it with three reference queries (prefab $\rightarrow$ mounted Animator $\rightarrow$ AnimatorController $\rightarrow$ clip) at 38,901 tokens and 10.4\,s; the exploration arm reassembled the same chain with 15 file operations, 194,486 tokens, and 176\,s ($5.0\times$ and $16.9\times$). The decisive hop crosses two GUID references that exist nowhere in C\# code.

\balance
\section{Related Work}\label{sec:related}

GUID-based asset analysis in Unity is long-established: Unity's scripting interface \texttt{AssetDatabase} has exposed GUID lookup for over a decade~\cite{unity-assetdatabase}, and community scanners have produced dependency reports for years. For agent evaluation, SWE-bench~\cite{swe-bench} established repository-level benchmarks and GameDevBench~\cite{gamedevbench} covers Godot; to our knowledge, no Unity-specific benchmark existed, which motivates our benchmark artifact.

\section{Conclusion}\label{sec:conclusion}

Asset-heavy repositories are where general-purpose agent tooling quietly fails: the decisive facts live in GUIDs and serialized YAML that no code-only view can see. Unity Insight closes the cross-file code--asset gap in production, and our paired experiment suggests it does so at roughly half the cost of raw agentic exploration. It is already applied in production as the built-in code--asset indexing layer of Tuanjie Codely, the agent CLI shipped with Tuanjie Engine (Unity China), where its five query tools serve as the agent's native navigation surface for Unity projects. More broadly, the recipe---a persistent, read-only code--asset graph exposed through typed query tools---is not tied to Unity: any engine whose decisive facts live in serialized assets rather than code can adopt the same shape. We expect such asset-aware indexes to become as standard a piece of agent infrastructure in game repositories as language servers~\cite{lsp} already are in mainstream code.

\bibliographystyle{ACM-Reference-Format}
\bibliography{refs}

@misc{codely-cli,
  author       = {{Unity China}},
  title        = {{@unity-china/codely-cli}: the Codely agent command-line interface},
  howpublished = {npm package registry},
  note         = {Insight tooling first public in version 1.0.0-rc.43 (2026-07-08); \url{https://www.npmjs.com/package/@unity-china/codely-cli}},
  year         = {2026}
}

@misc{unity-assetdatabase,
  author       = {{Unity Technologies}},
  title        = {{AssetDatabase} scripting interface},
  howpublished = {Unity scripting API documentation},
  note         = {Available since 2012; \url{https://docs.unity3d.com/ScriptReference/AssetDatabase.html}},
  year         = {2012}
}

@inproceedings{swe-bench,
  author       = {Jimenez, Carlos E. and Yang, John and Wettig, Alexander and Yao, Shunyu and Pei, Kexin and Press, Ofir and Narasimhan, Karthik},
  title        = {{SWE-bench}: Can language models resolve real-world {GitHub} issues?},
  booktitle    = {International Conference on Learning Representations (ICLR)},
  year         = {2024}
}

@inproceedings{gamedevbench,
  author       = {Chi, Wayne and Fang, Yixiong and Yayavaram, Arnav and Yayavaram, Siddharth and Karten, Seth and Wei, Qiuhong Anna and Chen, Runkun and Wang, Alexander and Chen, Valerie and Talwalkar, Ameet and Donahue, Chris},
  title        = {{GameDevBench}: Evaluating Agentic Capabilities Through Game Development},
  booktitle    = {International Conference on Machine Learning (ICML)},
  year         = {2026},
  note         = {arXiv:2602.11103; Godot-based benchmark}
}

@misc{lsp,
  author       = {Microsoft},
  title        = {Language {S}erver {P}rotocol specification},
  howpublished = {\url{https://microsoft.github.io/language-server-protocol/}},
  year         = {2023}
}

@misc{insight-docs,
  author       = {{Unity China}},
  title        = {Unity {I}nsight feature guide},
  howpublished = {Codely documentation, first archived 2026-09-02},
  note         = {\url{https://codely-docs.tuanjie.cn/features-introduction/unity-insight-guide}},
  year         = {2026}
}

@misc{redrunner,
  author       = {{BayatGames}},
  title        = {{RedRunner}: open-source 2D platformer game for Unity},
  howpublished = {GitHub repository, MIT License},
  note         = {\url{https://github.com/BayatGames/RedRunner}, accessed 2026-09-09},
  year         = {2017}
}

\end{document}